\documentclass[conference]{IEEEtran}
\IEEEoverridecommandlockouts
\usepackage{cite}
\usepackage{amsmath,amssymb,amsfonts}
\usepackage{algorithmic}
\usepackage{graphicx}
\usepackage{textcomp}
\usepackage{xcolor}
\usepackage{float}
\usepackage{listings}
\usepackage{courier}
\usepackage{hyperref}

\usepackage{dblfloatfix} % allows [!b] or [!h] for figure*

\graphicspath{{figs/}} % look for images in figs/

\def\BibTeX{{\rm B\kern-.05em{\sc i\kern-.025em b}\kern-.08em
    T\kern-.1667em\lower.7ex\hbox{E}\kern-.125emX}}
\begin{document}

\title{Accuracy and Efficiency Trade-Offs in LLM-Based Malware Detection and Explanation: A Comparative Study of Parameter Tuning vs. Full Fine-Tuning\\}

\author{\IEEEauthorblockN{ Stephen C. Gravereaux}
\IEEEauthorblockA{\textit{Department of Cybersecurity} \\
\textit{University at Albany - State University of New York}\\
Albany, NY, USA \\
sgravereaux@albany.edu}
\and
\IEEEauthorblockN{ Sheikh Rabiul Islam}
\IEEEauthorblockA{\textit{Department of Cybersecurity} \\
\textit{University at Albany - State University of New York}\\
Albany, NY, USA \\
sislam7@albany.edu}

}

\maketitle

\begin{abstract}
This study examines whether Low-Rank Adaptation (LoRA) fine-tuned Large Language Models (LLMs) can approximate the performance of fully fine-tuned models in generating human-interpretable decisions and explanations for malware classification. Achieving trustworthy malware detection, particularly when LLMs are involved, remains a significant challenge.
We developed an evaluation framework using Bilingual Evaluation Understudy (BLEU), Recall-Oriented Understudy for Gisting Evaluation (ROUGE), and Semantic Similarity Metrics to benchmark explanation quality across five LoRA configurations and a fully fine-tuned baseline. Results indicate that full fine-tuning achieves the highest overall scores, with BLEU and ROUGE improvements of up to 10\% over LoRA variants. However, mid-range LoRA models deliver competitive performance—exceeding full fine-tuning on two metrics—while reducing model size by approximately 81\% and training time by over 80\% on a LoRA model with 15.5\% trainable parameters.
These findings demonstrate that LoRA offers a practical balance of interpretability and resource efficiency, enabling deployment in resource-constrained environments without sacrificing explanation quality. By providing feature-driven natural language explanations for malware classifications, this approach enhances transparency, analyst confidence, and operational scalability in malware detection systems.

\end{abstract}


\section{Introduction}
Modern AI-based malware detection systems often lack trustworthiness, particularly when LLMs are involved, limiting analysts’ ability to validate automated decisions and improve detection strategies. As cyber threats grow in complexity, the need for transparent and explainable malware classification has become an urgent priority. Recent advances in Large Language Models (LLMs) offer a promising avenue for generating human-readable explanations of classification decisions. However, fine-tuning LLMs for domain-specific interpretability introduces significant computational challenges. Full-parameter fine-tuning achieves high accuracy but demands substantial resources, making it impractical for many operational environments. Low-Rank Adaptation (LoRA) has emerged as an efficient alternative, reducing memory and storage requirements while preserving much of the base model’s knowledge. Yet, the trade-offs between explanation quality and resource efficiency in LoRA versus full fine-tuning remain underexplored in the context of malware detection.

To address this gap, we propose a multistage pipeline that transforms EMBER malware classification outputs into human-interpretable explanations using LLM fine-tuning. Our approach integrates SHAP-based feature importance analysis to generate structured ground-truth explanations, which serve as training targets for both LoRA and fully fine-tuned models. We evaluate five LoRA configurations against a fully fine-tuned baseline using BLEU, ROUGE, and semantic similarity metrics to quantify explanation quality. Additionally, we analyze computational efficiency in terms of training time and model size.

Our \textbf{contributions} are fourfold:
\begin{itemize}

    \item A comparative study of LoRA and full fine-tuning approaches for LLM-based malware explanation generation.
    \item A standardized evaluation framework for assessing explanation quality using BLEU, ROUGE, and semantic similarity.
    \item Empirical insights into accuracy-efficiency trade-offs, identifying an ``optimal point'' in LoRA configurations that balances performance and resource savings.
    \item Practical implications for deployment, enabling interpretable malware detection in resource-constrained environments.
\end{itemize}
By demonstrating that LoRA can deliver competitive explanation quality with up to 81\% reduction in model size and 80\% faster training based off of the 15.5\% LoRA model, this work advances the feasibility of transparent, scalable malware detection systems.

\section{Background Study}

Explainable Artificial Intelligence (XAI) in cybersecurity has gained significant attention as machine learning models become increasingly opaque. Anderson et al. [1] demonstrated the vulnerability of machine learning-based malware detection systems to adversarial attacks, showing that adversarial examples can fool state-of-the-art detectors while preserving malicious functionality. This finding challenges the assumption that high accuracy alone is sufficient for deployment and emphasizes the need for interpretable models that allow analysts to understand and validate decisions. Without interpretability, defenders cannot distinguish legitimate detections from false positives caused by adversarial manipulation, nor identify weaknesses that attackers might exploit. They utilized an open-source OpenAI Gym environment in which a deep Q-learning agent repeatedly mutates a PE file using functionality-preserving actions, awarding a positive score when attack evasion succeeds. This also reinforces why explanations should reveal which features contribute to a model’s decision if we want to prevent RL agents from steering samples past it.

Recent advances in AI for malware analysis have focused on improving both detection accuracy and interpretability. Djenna et al. [2] provided a comprehensive survey of AI-based malware detection and mitigation techniques, highlighting the importance of incorporating domain knowledge into explainable AI. This is framed around understanding which threats emerge, which vulnerabilities they exploit, what vectors they use, and which operating modes violate security. With this understanding, it becomes easier to map them to intrusion detection and cyber threat intelligence systems, thereby helping to reduce the impact of potential malicious actors. Similarly, Islam et al. [18, 19, 20] infused domain knowledge for better interpretation of attack detection results.  

Faruk et al. [4] reinforced this by showing that understanding which features drive classification decisions is essential for improving models and providing actionable insights. When analysts can validate a model’s reasoning, they are better equipped to identify false positives and understand what makes a file malicious or benign. They perform feature analysis before classification by employing Fisher scores, information gain, and uncertainty symmetry on malicious samples sourced from Kaggle. This step is critical because accurate classification requires a thorough understanding of the underlying features that influence sample behavior. For this reason, integrating SHAP into the methodological pipeline is essential to ensure interpretability and transparency. Furthermore, SHAP-guided LoRA explanations should be evaluated for consistency across lightweight and heavy adapters to confirm or refute alignment in reasoning between models.

Emerging research explores whether Large Language Models (LLMs) can enhance interpretability in cybersecurity. Patsakis et al. [3] demonstrated that LLMs can reverse-engineer obfuscated malware and explain its behavior in plain language, even when code is encrypted or polymorphic. Similarly, Guo et al. [16] and Al-Karaki et al. [17] reviewed LLM applications in malware detection, noting their potential for static and dynamic analysis. However, most existing work focuses on code-level interpretation rather than explaining engineered features extracted by static analysis tools. This leaves an open question: Can LLMs explain malware classifications based on features like those in EMBER?

The EMBER dataset provides an excellent foundation for addressing this question. Researchers have extensively analyzed which of its 2,381 features matter most for detection. Oyama et al. [5] found that import table analysis and section entropy are strong indicators of malicious behavior, while Sandor et al. [6] showed that understanding feature correlations improves interpretability. These insights guide our approach: we use LightGBM combined with SHAP to identify feature importance and fine-tune LLMs to generate natural language explanations that reflect the reasoning behind EMBER scores.

Despite progress, a critical gap remains: techniques like SHAP provide numerical feature importance but fail to deliver clear, human-readable explanations for static feature-based classifications. Our work addresses this gap by integrating SHAP-based feature importance with LLM fine-tuning strategies (LoRA vs. full fine-tuning) to generate natural language explanations for EMBER-based malware classifications. This approach enhances interpretability while balancing accuracy and efficiency, enabling practical deployment in resource-constrained environments.

\section{Methodology}
Our methodology leverages the EMBER dataset to train five LLM models with varying percentages of trainable parameters and compares their performance against a fully fine-tuned model. The process begins by training models on the EMBER dataset, transforming raw malware detection outputs into human-interpretable explanations.
The architecture consists of four primary components:
\begin{itemize}
\item  EMBER feature extraction and classification
\item Feature importance ranking
\item LLM training and fine-tuning
\item Automated explanation generation using LoRA and full-parameter fine-tuning
\end{itemize}

By varying LoRA’s rank r—which controls the percentage of trainable parameters—we compare outputs against a fully fine-tuned model.

\begin{figure*}
    \centering
    \includegraphics[width=.8\linewidth]{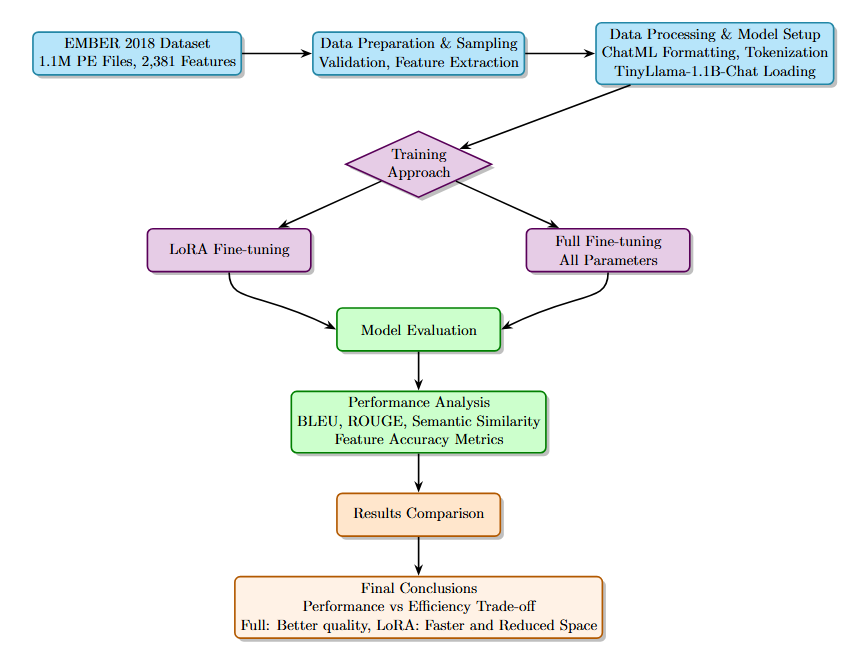}
    \caption{Pipeline Flowchart}
    \label{fig:pipeline}
\end{figure*}

\begin{figure}
    \centering
    \includegraphics[width=.5\linewidth]{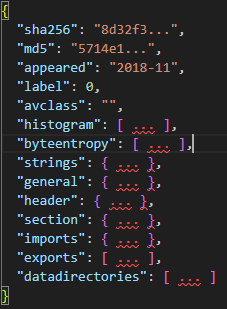}
    \caption{Truncated PE file extracted from the JSONL test features}
    \label{fig:pe}
\end{figure}

\textbf{Data Preparation}: Our pipeline (See Fig.\ref{fig:pipeline}) starts by downloading the EMBER 2018 dataset from the official repository. Feature extraction involves loading each PE file, applying static analysis techniques to extract features such as import tables, section headers, string content, and metadata (See Fig. \ref{fig:pe}), and converting these features into standardized 2,381-dimensional numerical vectors using the EMBER LightGBM model.

\begin{figure*}
    \centering
    \begin{minipage}{0.48\textwidth}
        \centering
        \includegraphics[width=\linewidth]{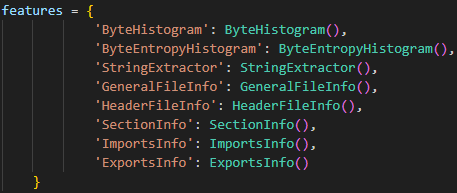}
        \caption{Located within the official EMBER dataset, lines 496–510 show the feature list.}
        \label{fig:ember-features1}
    \end{minipage}
    \hfill
    \begin{minipage}{0.48\textwidth}
        \centering
        \includegraphics[width=\linewidth]{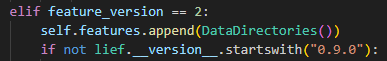}
        \caption{The 9th category is added on line 526 only for feature version 2 (the current version).}
        \label{fig:ember-features2}
    \end{minipage}
\end{figure*}

From the EMBER dataset, we selected 1,050 representative samples: 1,000 for training and 50 for testing and validation. Sampling maintained a balanced distribution to ensure unbiased training. Both LoRA and full-parameter models were trained on the same data. Samples included diverse malware families (e.g., xrat, ramnit, sality, zbot, emotet, ursnif, lethic) and benign software types (system utilities, applications, drivers) to maximize coverage of PE characteristics. The test set consisted of 50 samples (25 malware, 25 benign), meeting criteria such as verified ground truth labels, 2,381-dimensional feature vectors, and varied file sizes. For evaluation, 5 samples (3 malware, 2 benign) were selected as a focused test set.
\begin{figure*}
    \centering
    \includegraphics[width=1.0\linewidth]{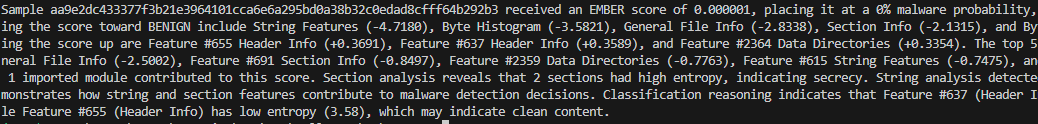}
    \caption{Full parameter output on EMBER PE sample}
    \label{fig:full_parameter}
\end{figure*}

\begin{figure}
    \centering
    \includegraphics[width=1\linewidth]{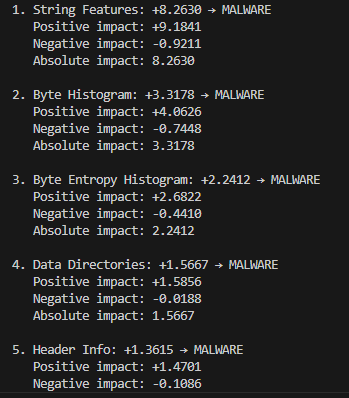}
    \caption{Shap results for top 5 features contributing to malware classification decision from EMBER PE file}
    \label{fig:shap}
\end{figure}

\textbf{Training Pipeline}: The pipeline converts raw features into input-output pairs for LLM training using SHAP (SHapley Additive exPlanations). SHAP values quantify feature importance by calculating each feature’s contribution to the model’s prediction. We use SHAP’s TreeExplainer on the EMBER LightGBM model, producing 2,381 SHAP values aggregated into nine critical feature groups (See Fig. \ref{fig:ember-features1} and \ref{fig:ember-features2}): suspicious API's (high-risk function calls), entropy analysis (packing and obfuscation indicators), import patterns (library dependencies and API usage), section characteristics (code section analysis), file size indicators (size-based classification factors), string analysis (embedded content patterns), PE structure (file format metadata), and malware family identification (family-specific signatures). Top contributing feature groups are then fed into the models to generate natural language explanations describing their influence on EMBER’s classification (See Fig. \ref{fig:shap}).

\textbf{Input and Output Formats}: The input format for LLM training includes the EMBER score (0.0–1.0), top five shap feature groups, binary classification label (malware/benign), and a structured prompt requesting explanation on the classification. The output format contains SHAP-enhanced explanations detailing EMBER’s confidence level, the top five feature groups ranked by absolute SHAP contribution, their directional influence (toward malware or benign), and impact level. This SHAP-based truth serves as the target for both LoRA and full-parameter fine-tuning, ensuring the LLM replicates the EMBER classification model's reasoning process. A unified prompt template using ChatML format (system instructions, user queries, assistant responses) ensures consistency across both fine-tuning approaches. Outputs generated by the full-parameter model (See Fig. \ref{fig:full_parameter}) explain each of the top five SHAP features in detail in the generated explanation.

\textbf{Evaluation}: We assess explanation quality using BLEU, ROUGE, and semantic similarity scores: \textbf{ BLEU} measures how much of the generated text is present in the reference ground truth. It does this by comparing its word sequences, known as n-grams. \textbf{ROUGE} focuses on how much of the reference information is captured in the explanation. Since both of these are text comparisons, we additionally use \textbf{ semantic similarity} scores using embedding cosine similarity that assess meaning alignment, even when phrasing differs.

\section{Data}
The EMBER 2018 dataset is a comprehensive benchmark for static malware detection, containing 1,000,000 PE (Portable Executable) files with verified labels. The dataset is structured into 600,000 training samples, 200,000 validation samples, and 200,000 test samples. Each PE file includes exactly 2,381 static features organized into nine categories (See Fig. \ref{fig:pe}, \ref{fig:ember-features1}, and \ref{fig:ember-features2}):

\begin{itemize}
    \item \textbf{Import analysis features}: 1,280 dimensions covering API function calls, library dependencies, and import characteristics
    \item \textbf{Byte histogram features}: 256 dimensions representing byte frequency distributions
    \item \textbf{Byte entropy histogram features}: 256 dimensions capturing entropy patterns
    \item \textbf{Section analysis features}: 255 dimensions including entropy calculations, section sizes, and structural indicators
    \item \textbf{Export analysis features}: 128 dimensions covering exported function information
    \item \textbf{String analysis features}: 104 dimensions capturing embedded strings, character distributions, and encoding patterns
    \item \textbf{Header analysis features}: 62 dimensions covering PE header characteristics and structural elements
    \item \textbf{Data directory features}: 30 dimensions covering resource information and directory structures
    \item \textbf{General file information}: 10 dimensions including file size, compilation timestamps, and metadata
\end{itemize}

The dataset includes diverse malware families such as xrat, ramnit, sality, zbot, emotet, ursnif, and lethic, representing real-world variants. EMBER provides these features in a safe format using JSONL files. Fig. \ref{fig:pe} shows an extracted PE file sample from the EMBER test features JSONL. It is truncated to show only the fields relevant to this analysis.
This sample (Fig.\ref{fig:pe}) is benign, indicated by its label value of 0. It includes SHA256 and MD5 hashes. The histogram field contains 256 integer values representing byte frequencies, which can reveal patterns since malware often exhibits unusual distributions due to packing. The strings object summarizes embedded text; this sample has 321 strings with an average length of 13.12 characters. The paths, URLs, and registry fields are all zero, indicating no embedded file paths. The general object provides metadata such as disk and memory size, exports, imports, and API functions. The sections object describes internal sections, while the imports object lists Windows APIs used—critical for malware detection.
The exports array lists four exported functions, analyzed by the LightGBM model. For this sample, the model would produce a low malware score because the file exhibits benign characteristics: normal entropy levels, reasonable imports, legitimate section names, and no suspicious API combinations.
\section{Experiments}

Several open-source tools and libraries were utilized for various components of the research. For LLM fine-tuning, we utilized the Hugging Face Transformers library with AutoTokenizer and AutoModelForCausalLM for model loading and inference, the PEFT library for LoRA implementation using LoraConfig and get\_peft\_model functions, and standard PyTorch training loops with the AdamW optimizer and learning rate scheduling.
For SHAP, we employed the SHAP library’s TreeExplainer to calculate feature importance from the EMBER LightGBM model. This was used in conjunction with background sampling and aggregation across the nine EMBER feature groups.

The model evaluation uses NLTK’s sentence BLEU function for BLEU score calculation with smoothing methods, the ROUGE score library for computing ROUGE-1, ROUGE-2, and ROUGE-L metrics, and Sentence-Transformers with the all-MiniLM-L6-v2 model for generating embeddings and calculating cosine similarity for semantic reasoning. All evaluations compare model-generated explanations against the SHAP ground truth explanations from the EMBER model.

Low Rank Adaptation (LoRA) fine-tuning was implemented using the PEFT library with
specific configuration parameters. The LoRA approach decomposes weight updates into low-rank matrices, allowing efficient adaptation while preserving the base model’s knowledge. We trained five LoRA variants with ranks of 16, 96, 256, 512, and 896 corresponding to 1.15 percent, 6.44 percent, 15.50 percent, 26.85 percent, and 39.11 percent of trainable parameters respectively. A higher rank indicates more trainable parameters. Each configuration used a scaling parameter (lora alpha) of 32 to control the magnitude of LoRA updates and a dropout probability of 0.1 to prevent overfitting in the LoRA layers. The adaptation targeted transformer components, including query, key, value, and output projection layers in the attention mechanism, as well as gate, up, and down projections in the feed-forward networks. 

The base TinyLlama-1.1B chat model was loaded using AutoModelForCasualLM
with a half-precision floating point (torch.float.16) to optimize memory usage. The training was performed using the AdamW optimizer with a learning rate of \(5 \times 10^{-5}\), a weight decay of 0.01 for regularization, and a batch size of one sample per iteration. The training schedule included one epoch with gradient accumulation over four steps to simulate larger batch sizes. Full Parameter fine-tuning was implemented using PyTorch training procedures of the TinyLlama1.1B LLM. The model was loaded with torch dtype=torch.float16 and
gradient checkpointing=true to efficiently manage memory usage. The training configuration included AdamW optimizer with a learning rate of \(5 \times 10^{-6}\). (lower than LoRA to prevent catastrophic forgetting), weight decay = 0.01, batch size = 1 (3070TI 8 GB GPU memory), num train epochs=1, gradient accumulation steps=8, warmup steps=5. The loss convergence was observed, with progression from 1.8982 initially to 0.6178 at the end.

To evaluate the effectiveness of different fine-tuning approaches. We conducted a comparison on 50 balanced test samples. The evaluation framework compares explanations generated by each LoRA model against those produced by the full-parameter model, assessing how closely LoRA models replicate full-parameter performance. For each of these samples, we use SHAP TreeExplainer to identify the top 5 feature groups, which are incorporated into the prompts to provide context for the LLMs. This was done the same way for creating the training data for the 1000 training samples. For each of the 50 test samples, we generated explanations using all six models (Five LoRA variants and one full parameter model) and calculated five metrics: BLEU score for n-gram precision, ROUGE-1 for uni-gram recall, ROUGE-2 for bi-gram recall, ROUGE-L for longest common subsequence, and semantic similarity. Due to repository size constraints and the large size of LLM variants, only essential code files will be shared on GitHub at https://github.com/StephenCGravereaux/EMBER-LLM-Comparison-Pipeline
\section{Results}

\begin{figure*}[!t]
    \centering

    \begin{minipage}{0.48\textwidth}
        \centering
        \includegraphics[width=\linewidth]{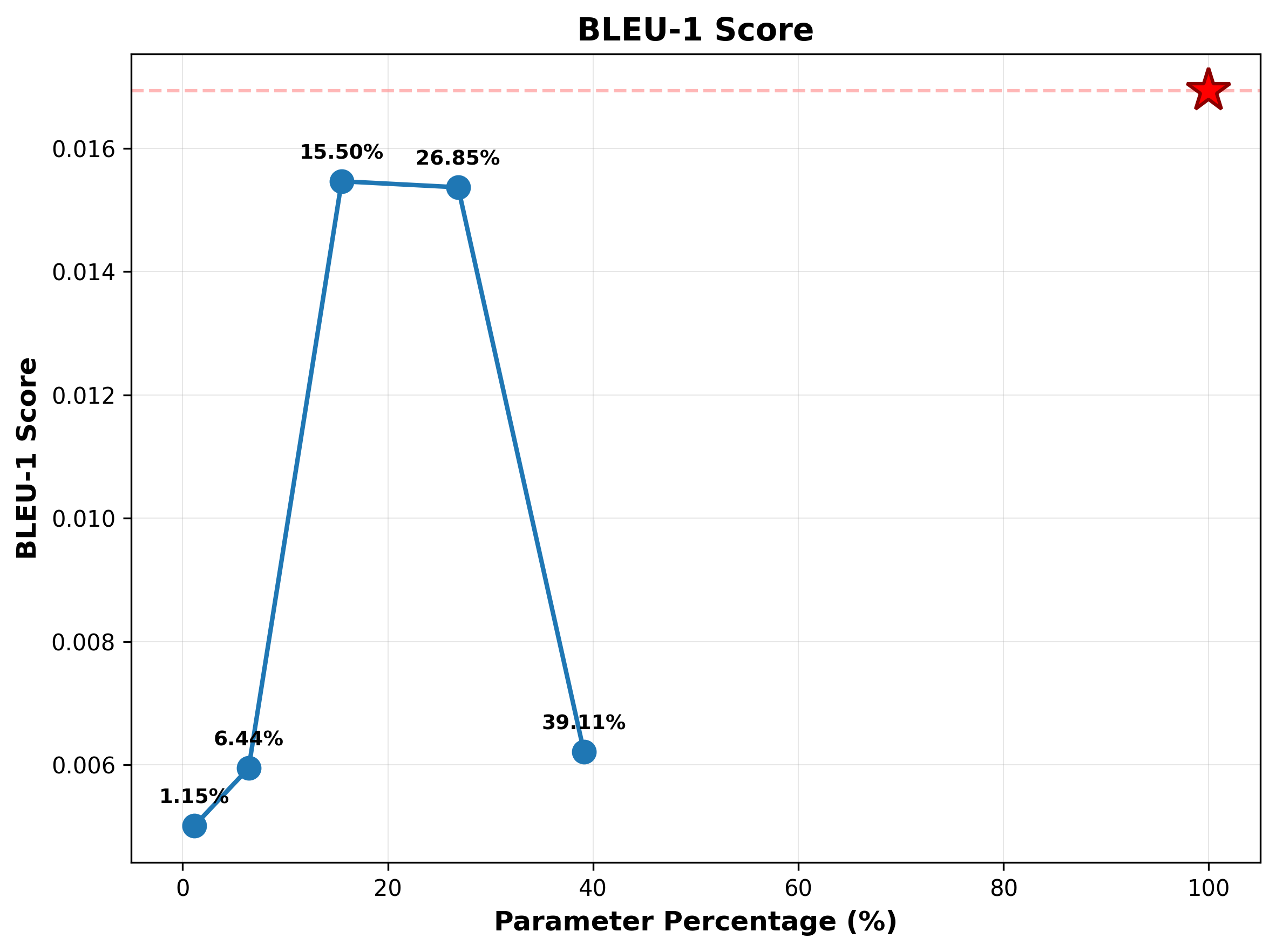}
        \caption{BLEU Score}
        \label{fig:bleu}
    \end{minipage}
    \hfill
    \begin{minipage}{0.48\textwidth}
        \centering
        \includegraphics[width=\linewidth]{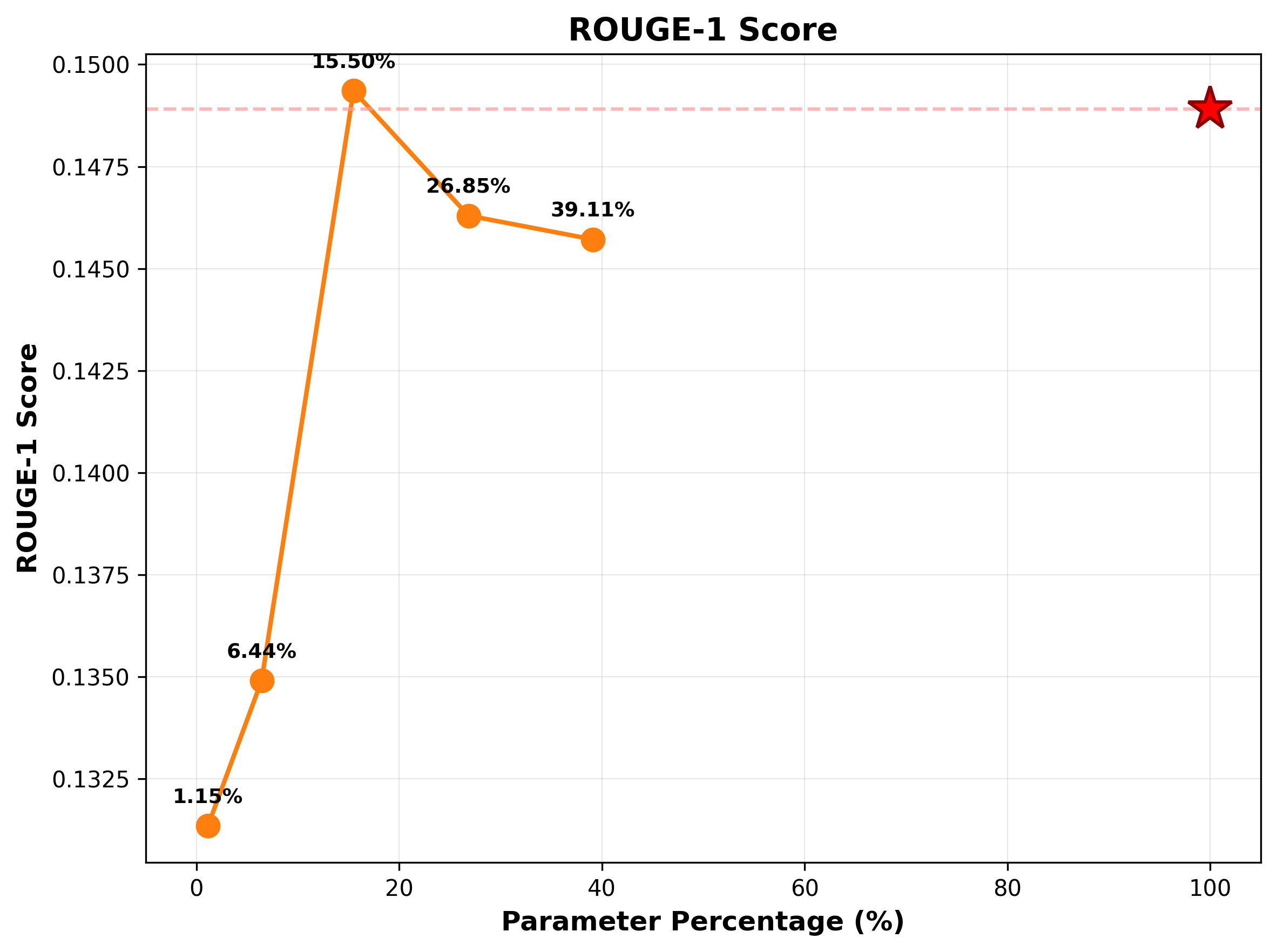}
        \caption{ROUGE-1 Score}
        \label{fig:rogue1}
    \end{minipage}

\end{figure*}

\begin{figure*}[!t]
    \centering

    \begin{minipage}{0.48\textwidth}
        \centering
        \includegraphics[width=\linewidth]{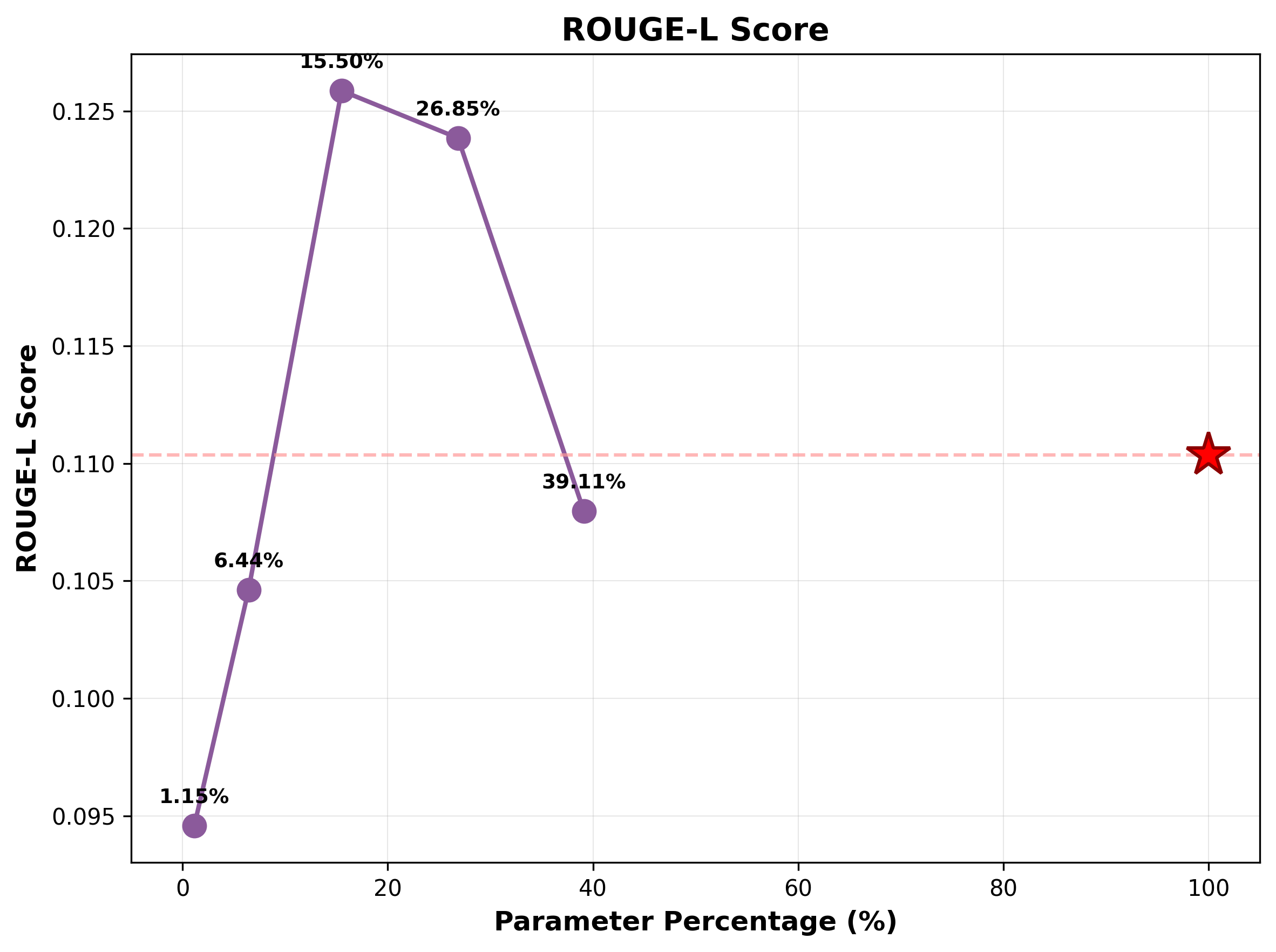}
        \caption{Rouge-L Scores}
        \label{fig:ROUGE-1}
    \end{minipage}
    \hfill
    \begin{minipage}{0.48\textwidth}
        \centering
        \includegraphics[width=\linewidth]{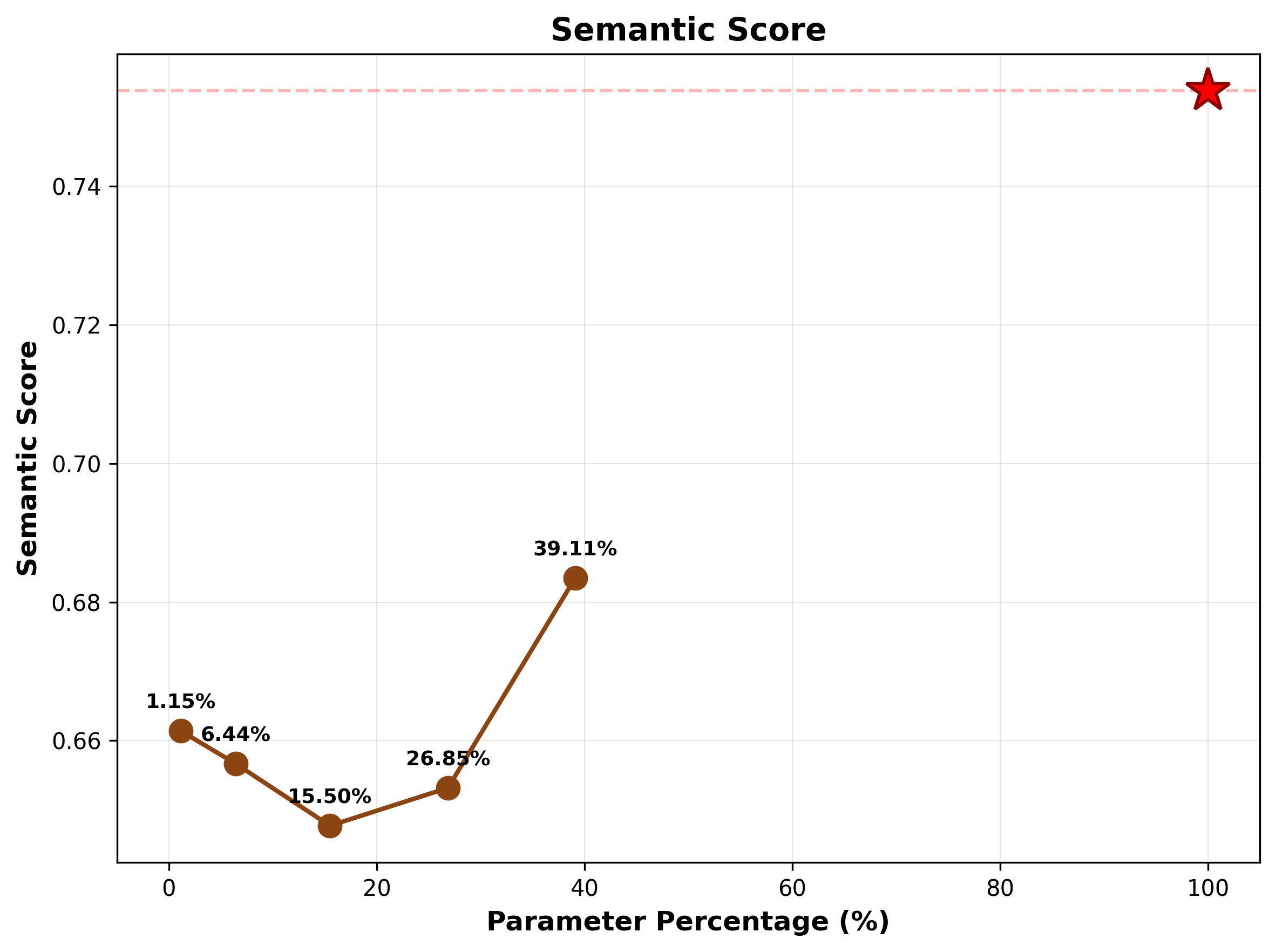}
        \caption{Semantic Scores}
        \label{fig:semantic}
    \end{minipage}

\end{figure*}
\begin{figure}
    \centering
    \includegraphics[width=1\linewidth]{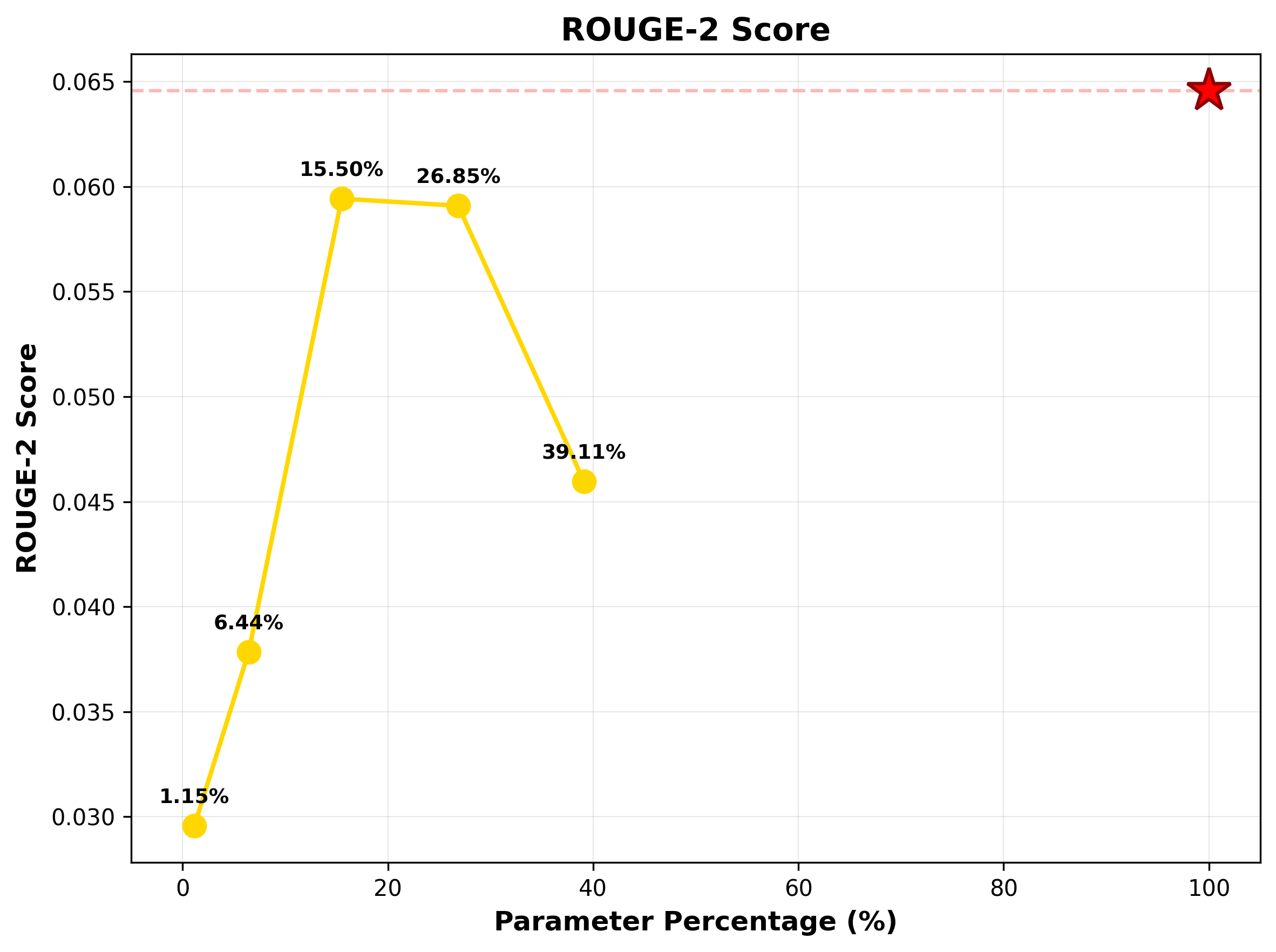}
    \caption{ROGUE-2 Score}
    \label{fig:ROUGE-2}
\end{figure}

\begin{figure*}[t]
    \centering
    \includegraphics[width=.8\linewidth]{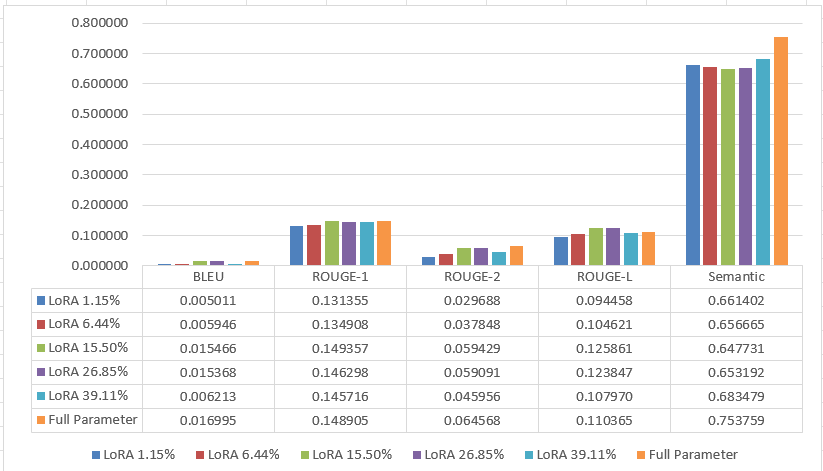}
    \caption{BLEU, ROUGE, and Semantic results}
    \label{fig:results}
\end{figure*}

The analysis of LoRA variants versus full fine-tuning reveals key insights into the effectiveness of different parameter training methodologies. Since each LLM produces 50 explanations and a corresponding score for each, we compute the average score across all 50 samples for each metric.
When comparing all six models, the full-parameter model achieved the highest scores across most metrics (See Fig. \ref{fig:results}). It scored a BLEU (See Fig. \ref{fig:bleu}) of 0.0169, ROUGE-1(See Fig. \ref{fig:rogue1}) of 0.1489, ROUGE-2 of 0.0646(See Fig. \ref{fig:ROUGE-2}, ROUGE- L (See Fig. \ref{fig:ROUGE-1}) of 0.1104, and semantic similarity (See Fig. \ref{fig:semantic}) of 0.7538. Among the LoRA variants, LoRA 15.50 percent (rank 256) showed strong overall performance with BLEU of 0.0155, ROUGE-1 of 0.1494 (exceeding Full Parameter), ROUGE-2 of 0.0594, ROUGE-L of 0.1259 (exceeding Full Parameter), and semantic similarity of 0.6477.

The relationship between parameter percentage and performance reveals a non-monotonic efficiency curve where updating percentages beyond a certain threshold may lead to diminishing returns and potential overfitting at higher ranks. The mid-range LoRA models consistently outperformed both smaller and larger variants, indicating an "optimal point" where the model has sufficient capacity to learn without overfitting the training data. LoRA at 39.11 percent showed the highest semantic similarity among the LoRA models (See Fig. \ref{fig:semantic}). This suggests that while the model didn't closely replicate the formatted output style, it still produced better explanations relative to contextual meaning. With BLEU scores remaining relatively low across all models, which can be attributed to the technical diversity of vocabulary in malware explanations, where a lot of valid phrasings exist. The ROUGE metrics are also low but have slightly higher scores, indicating some vocabulary overlap. Semantic similarity emerged as the most discriminative metric with higher scores across the board, demonstrating that all models successfully capture the conceptual meaning of the classification statements despite vocabulary variations.

\subsection{File size}
\begin{figure*}[t]
    \centering
    \label{fig:training_time}
    \includegraphics[width=.8\linewidth]{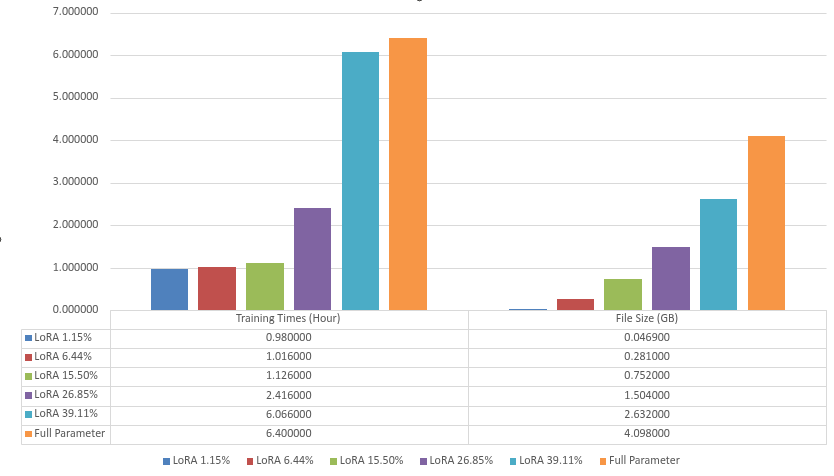}
    \caption{Training time and file size}
    
\end{figure*}

Statistical data was collected for training times across each model, along with the size of the model itself. This is valuable information as it reflects the computational cost and performance of each model relative to its size. The relationship between LoRA rank (r) and model size follows a linear growth. For each adapted module, LoRA introduces two low-rank trainable matrices: a down size (d x r) and an up projection of size (r x z), where d is the original hidden dimension and r is the chosen rank. Z is the output dimensionality of the target weight. TinyLlama-1.1B applies LoRA to seven modules across twenty-two transformer layers. This corresponds to 154 injection points. Higher LoRA ranks increase the amount of weights and thus lead to a larger file size. Full fine-tuning, on the other hand, updates all parameters in the model rather than only the low-rank adapter matrices. Consequently, the optimizer must maintain and update the entire set of tensors in the base model. This explains why a full-finetuned model is significantly larger in size compared to any LoRA-based configuration. The file sizes confirm this relationship from each of the models created.LoRA r=16 produced a 48.23 MB adapter file. LoRA R96 resulted in a 288.79 MB file which demonstrates a linear growth (See Fig. \ref{fig:training_time}). As the rank increases, the model sizes reaches 770 MB, 1,540 MB, and 2,695 MB for r=256, r=512, and r=896, respectively. In comparison, the full parameter model occupied 4,196 MB, representing a substantial reduction in storage requirements.

This substantial reduction highlights the practical advantages of LoRA models for optimizing storage. With smaller file sizes, models can be loaded faster and distributed more efficiently. For example, there is an 81.65\% decrease in file size between the fully fine-tuned model and the r=256r = 256r=256 configuration (with 15.5\% of parameters trained).

\subsection{Training time}
During training, several computational bottlenecks were observed, particularly for higher-rank LoRA configurations. The most significant bottleneck was GPU memory constraints on the NVIDIA RTX 3070 Ti (8 GB VRAM). Lower rank models (r-16, r=96, r=256) trained efficiently with batch size 8 and no gradient checkpointing. These models were trained in times of 59 minutes, 61 minutes, and 73 minutes, respectively. LoRA r=512 model and LoRA r=812 (26 percent and 39 percent) took 145 minutes and 364 minutes. These times are longer, but the full parameter model was the longest, taking 384 minutes. 
The LoRA r=896 model required more aggressive memory optimizations, such as a reduced batch size of 2, and a gradient accumulation over 4 steps, enabling checkpointing. Gradient check pointing trades computation for memory by recomputing intermediate activations during the backward pass rather than storing them. This technique doubles the computational time, which explains why it almost took as long as the full parameter model despite having fewer updated parameters.
   
\subsection{Natural language explanations}
As a byproduct of training these models, we have provided deeper insight into the underlying EMBER malware classification system. While the system originally outputs only a numeric score between 0 and 1 to indicate benign or malicious classification, we have added an interpretability layer that explains the classifier’s reasoning to analysts. Although the EMBER classifier operates with strong accuracy, it lacks the ability to justify its decisions. This added feature enhances transparency and helps analysts understand the underlying decision-making process that would otherwise remain opaque.

\section{Conclusion}

This research demonstrates that Large Language Models (LLMs) can be effectively fine-tuned to generate human-interpretable decisions for malware classifications. Our findings show that the fully fine-tuned model achieved the highest scores across most evaluation metrics, confirming its superior explanation quality. However, among the five LoRA variants tested, mid-range configurations (e.g., ~15.5\% trainable parameters) provided the best balance between performance and efficiency, even surpassing the full-parameter model on two metrics.
The ability to produce natural language explanations for malware classifications has significant practical implications for malware detection systems. Security analysts can now access detailed reasoning behind classification decisions, improving trust and enabling better validation of automated systems. 

Furthermore, LoRA-based models offer a compelling solution for resource-constrained environments, reducing model size by up to 81\% and training time by over 80\% while maintaining competitive explanation quality.
Based on these results, we recommend full-parameter fine-tuning when explanation quality is critical and computational resources are abundant. For lightweight deployments or scenarios requiring rapid model transfer and training, LoRA with approximately 15.5\% trainable parameters offers the most favorable trade-off between accuracy and efficiency.

Future work will focus on scaling this approach to larger LLMs beyond TinyLlama-1.1B. Comparative evaluations using BLEU, ROUGE, and semantic similarity across higher-capacity models may reveal new performance trends and help identify an “optimal point” for different LoRA configurations relative to full fine-tuning. Such investigations could further advance the development of interpretable, resource-efficient malware detection systems. Additionally, using a larger training dataset instead of the current subset of 1,000 samples might provide more robust results. In addition, expanding the training data will improve model generalization and reduce variance as well as overfitting. It will also enable the model to learn more discriminative feature patterns in low-frequency malware classes that are underrepresented in the current subset.

\section*{Acknowledgments}
This research was supported by the Minerva Center Innovation Funding for Research \& Creative Endeavors at the University at Albany. The authors sincerely appreciate this support, which was instrumental in the successful completion of this research.

\end{document}